\documentclass{article}
\usepackage{graphicx} 
\usepackage[
colorlinks=false
]{hyperref}
\usepackage[
backend=biber,
style=alphabetic,
]{biblatex}
\usepackage{authblk}
\usepackage{amsmath}
\usepackage[colorinlistoftodos]{todonotes}

\title{Detachment scalings derived from 1D scrape-off-layer simulations}

\author[1]{Thomas Body}
\author[1]{Thomas Eich}
\author[1]{Adam Kuang}
\author[1]{Thomas Looby}
\author[2]{Mike Kryjak}
\author[3]{Ben Dudson}
\author[1]{Matt Reinke}
\affil[1]{Commonwealth Fusion Systems}
\affil[2]{School of Physics, Engineering and Technology University of York, Heslington, York YO10 5DD, UK}
\affil[3]{Lawrence Livermore National Laboratory, 7000 East Avenue, Livermore CA 94550, USA}

\date{April 2024}

\begin{document}

\maketitle
\section{Abstract}

Fusion power plants will require detachment to mitigate sputtering and keep divertor heat fluxes at tolerable levels. Controlling detachment on these devices may require the use of real-time scrape-off-layer modeling to complement the limited set of available diagnostics. In this work, we use the configurable Hermes-3 edge modeling framework to perform time-dependent, fixed-fraction-impurity 1D detachment simulations. Although currently far from real-time, these simulations are used to investigate time-dependent effects and the minimum physics set required for control-relevant modeling. We show that these simulations reproduce the expected rollover of the target ion flux — a typical characteristic of detachment onset. We also perform scans of the input heat flux and impurity concentration and show that the steady-state results closely match the scalings predicted by the 0D time-independent Lengyel-Goedheer model. This allows us to indirectly compare to SOLPS simulations, which find a similar scaling but a lower value for the impurity concentration required for detachment for given upstream conditions. We use this result to suggest a series of improvements for the Hermes simulations, and finally show simulations demonstrating the impact of time-dependence.

\section{Introduction}

Detachment will be required in fusion power plants such as ARC or DEMO to limit erosion of the divertor targets and keep heat fluxes within tolerable limits \cite{Leonard2018-xi,Kuang2018-xt}. By detachment, we mean a pronounced drop of both the power and particle fluxes to the divertor target due to impurity radiation, plasma-neutral interactions and volumetric recombination within a flux-tube\footnote{A loss of both particle and power fluxes is termed  `ultimate detachment'  in Ref \cite{Krasheninnikov2017-ay}, and is contrasted to `partial' and `full' detachment which refer to cross-field profiles at the divertor target} \cite{Krasheninnikov2017-ay}. Under these conditions, the divertor target cools and a `detachment front' moves into the divertor volume \cite{Lipschultz2016-iz,Krasheninnikov1999-lo} towards the X-point (smoothly in some experiments and as a fast bifurcation in others \cite{Krasheninnikov2017-ay}). Once the detachment front reaches the X-point, it can degrade core performance by cooling the confined-region plasma and by reducing impurity screening \cite{Lipschultz2016-iz}. The detachment front can be stabilized at the X-point (referred to as an `X-point radiator') or can become unstable (referred to as a `MARFE') and trigger a density-limit disruption \cite{Bernert2023-gp}.

Several devices have demonstrated control of radiative power dissipation via divertor target measurements. On ASDEX Upgrade the divertor shunt current and a subset of bolometry chords is used to control nitrogen and argon seeding in `double radiation feedback' \cite{Kallenbach2012-wa}. On Alcator C-MOD surface thermocouples were used estimate and actively control the surface heat flux \cite{Brunner2016-pg}. On JET target $I_{sat}$ measurements were used to maintain divertor detachment during strike-point sweeping. More recently, emission fronts have been successfully controlled between the divertor strike-point and the X-point. On TCV, the carbon-III emission front (measured by the MANTIS multispectral imaging system \cite{Perek2019-av}) was successfully feedback-controlled \cite{Ravensbergen2020-kg,Ravensbergen2021-wj}. On MAST-U, a similar method to ref \cite{Ravensbergen2021-wj} was used to control the molecular-deuterium Fulcher emission front \cite{Kool2024-fw,Verhaegh2023-ei} (measured by the MWI multispectral imaging system \cite{Feng2021-lj}) was controlled. Control of the X-point radiator position has also been successfully demonstrated on multiple machines \cite{Bernert2023-gp}.
Looking ahead to fusion power plants, target measurements may not be sufficient to protect the divertor targets since -- due to the high attached heat fluxes -- significant damage may occur between detecting reattachment with target measurements and regaining control. The X-point radiator by contrast is attractive in terms of control, but it is not yet clear whether this regime has sufficient energy confinement, impurity screening and pumping to be used on a power plant. We therefore focus on techniques to control the detachment front (or an associated emission front) between the divertor target and X-point. Since neutron loading will damage many diagnostic systems such as multispectral imaging \cite{Raukema2024-jz}, we are exploring the use of real-time scrape-off-layer modeling to complement a reduced set of diagnostics for detachment control*.

The use of modeling for detachment control is still under development. Typical SOLPS-ITER simulations take days to months to run\footnote{The computational cost of SOLPS-ITER depends on machine parameters and on the use of higher fidelity physical effects such as drifts or kinetic neutrals.}, which is orders of magnitude slower than what would be needed for control\footnote{On ASDEX Upgrade and JET, reattachment timescales are in the range of $100ms$ to $1s$\cite{Henderson2024-bw}. Control algorithms would likely need to run in some fraction of this.}. To accelerate these simulations for real-time use, sparse regression-based dynamics models based on time-dependent SOLPS-ITER simulations \cite{Lore2023-pp} and surrogates based on steady-state SOLPS-ITER simulations\cite{Wiesen2024-ns} have been recently developed. Surrogate models for time-dependent 1D Braginskii scrape-off-layer models have also been developed -- including for UEDGE-1D \cite{Zhu2022-qq}, DIV1D \cite{Poels2023-ry} and Hermes-3 \cite{Holt2024-hi}. These fast 1D and 2D models are promising, although as of writing these models have not yet been used for detachment control on real experiments. Model-based detachment control to date has mainly relied on much simpler models, such as the use of the two-point-model to estimate the attached ion saturation current in KSTAR's detachment controller \cite{Eldon2022-ws}.

In this work, we compare simple detachment models to higher fidelity 1D Braginskii equations with the goal of identifying simple models which could be relevant for control. We first develop 1D fixed-fraction-impurity scrape-off-layer simulations using Hermes-3 \cite{Dudson2024-ru}, and show that these simulations qualitatively reproduce the expected $j_{sat}$ rollover as the simulations are pushed towards detachment. We then develop a PID feedback controller to hold the detachment front at a desired spatial location and use this to identify the equilibrium relationship between the input heat flux, the upstream density and the concentration of the radiating impurity. These relationships can be described by simple power laws, which allows for a direct comparison to the Lengyel-Goedheer and Spitzer-Harm models. This comparison shows that the steady-state physics present in the Hermes simulations matches the simpler models.

\section{Simulation setup}

\begin{figure}
    \centering
    \includegraphics[width=0.8\linewidth]{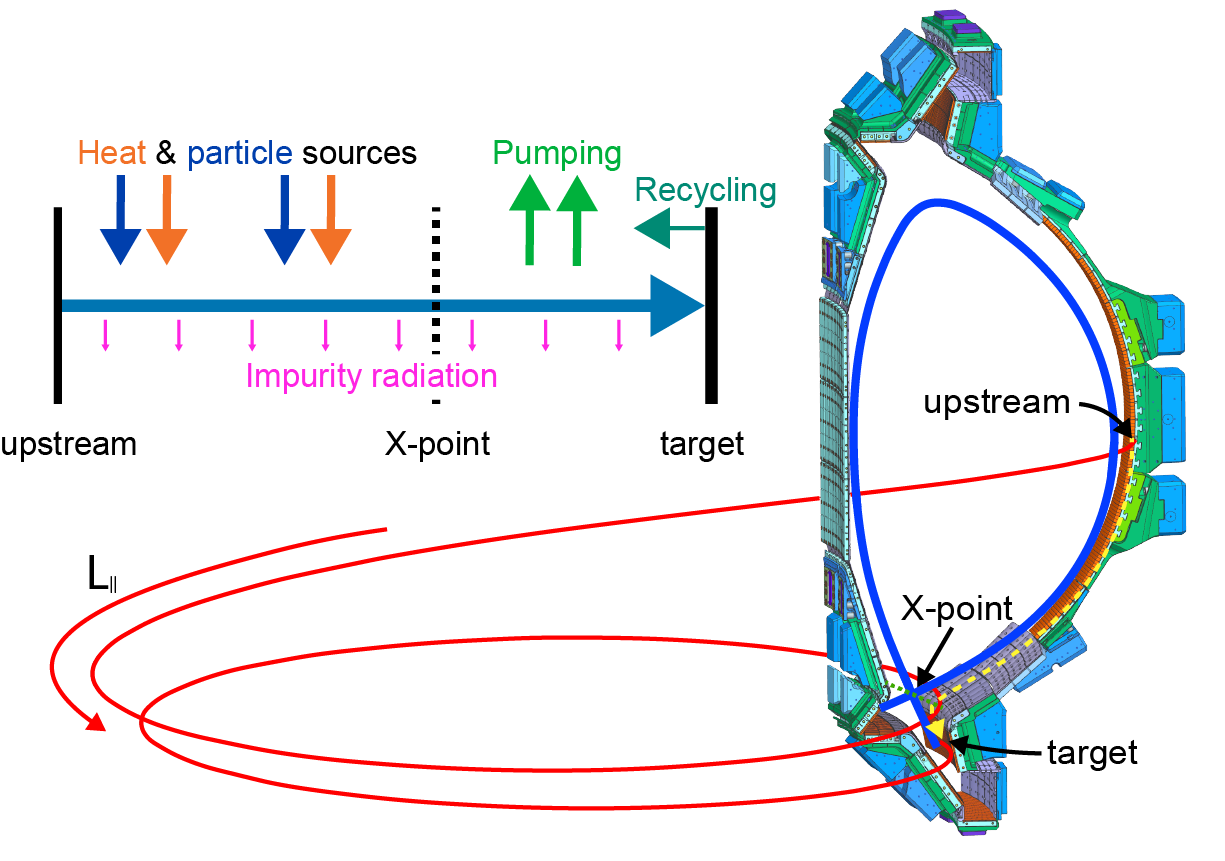}
    \caption{Schematic of the simulation setup. The 1D simulations follow a fieldline in the scrape-off-layer (typically $1\times\lambda_q$ from the separatrix) from the outboard midplane to the outer divertor target. The simulation domain is divided into the region above the X-point, where heat and particles are added, and below the X-point, where neutrals are pumped from the divertor. }
    \label{fig:hermes_setup}
\end{figure}

We perform Hermes 1D simulations along a single flux-tube in the scrape-off-layer, starting at the outboard midplane and following a field line to the low-field-side divertor target (as depicted in figure \ref{fig:hermes_setup}). The simulation domain is split at the point where the fieldline passes the X-point. Upstream of this point, particle and heat fluxes are added to the flux-tube, representing plasma transport from the confined region. The transport of energy and particles is then calculated according to the Braginskii equations \cite{Braginskii1965-qs} projected into 1D, as given in section \ref{sec:hermes_equations}.

\subsection{Model equations}\label{sec:hermes_equations}

In this work we treat only a single main-ion deuterium species with density $n_{d+}$, with a constant fixed-fraction neon impurity with density $n_{Ne} = c_{Ne} n_e$ where $n_{Ne}$ and $n_e$ are the neon and electron densities. We neglect the main ion dilution due to impurities, and therefore via quasineutrality assume that $n_e=n_{d+}$. We also neglect scrape-off-layer currents, equivalent to assuming that the electron and ion velocities $v_e$ and $v_{d+}$ are equal. Under this assumption, the parallel electric field reduces to $E_\parallel = -\frac{\partial_\parallel p_e}{e n_e}$ for $p_e$ the electron pressure ($\partial_\parallel$ indicates a gradient along the fieldline). These assumptions leave seven spatially-varying quantities which evolve in time — the deuterium ion and neutral densities $n_{d+}$ and $n_d$, the deuterium ion and neutral parallel velocities $v_{d+}$and $v_d$, and the deuterium ion, neutral and electron pressures $p_{d+}$, $p_d$ and $p_e$. For each species $s$, we define a temperature $T_s$ which is calculated from the species pressure and density $T_s = p_s/n_s$. 
The evolution of the deuterium ion and neutral densities are given by particle conservation equations
\begin{align}
\frac{\partial n_{d+}}{\partial t} 
    =& -\partial_\parallel\left( n_{d+}v_{d+} \right) + R_{iz} - R_{rec} + S_{n_{d+}}\\
    \frac{\partial}{\partial t}n_d = &
    -\partial_\parallel\left( n_d v_d \right) - R_{iz} + R_{rec} - \frac{n_d}{\tau_{pump}}
    +\partial_\parallel\left( D_d n_d \frac{\partial_\parallel p_d}{p_d} \right)\label{eq:neutral_continuity}
\end{align}
in terms of the particle advection, the ionization rate $R_{iz}=k_{iz}n_e n_d$, the recombination rate $R_{rec}=k_{rec}n_e n_d$, the external particle source rate $S_{n_{d+}}$ and the neutral particle divertor residence time $\tau_{pump}$. A pitch-angle dependent diffusion factor $D_n=\left(\frac{B}{B_{pol}}\right)^2\frac{T_n}{m_D \nu_{d}}$ (with the neutral collision frequency $\nu_d$ given by equation B10.c in ref \cite{Dudson2024-ru}) is used to roughly approximate the cross-field transport of neutrals. Each factor of pitch angle accounts for a crude mapping between perpendicular and parallel gradients, assuming that the plasma solution is constant in toroidal angle.
The evolution of the deuterium ion and neutral velocities are given by momentum conservation equations
\begin{align}
    \frac{\partial}{\partial t}\left( m_{d+} n_{d+} v_{d+} \right)
    =& -\partial_\parallel\left( m_{d+}n_{d+}v_{d+}\cdot v_{d+} \right) - \partial_\parallel p_{d+} + e n_{d+}E_\parallel\nonumber\\
    & +m_{d}v_d R_{iz} - m_{d+}v_{d+}R_{rec} + m_{d+}(v_d - v_{d+})R_{CX}\\
    \frac{\partial}{\partial t}\left( m_{d} n_{d} v_{d} \right)
    =& -\partial_\parallel\left( m_{d}n_{d}v_{d}\cdot v_{d} \right) - \partial_\parallel p_{d}\nonumber\\
     &-m_{d}v_d R_{iz} + m_{d}v_{d+}R_{rec} - m_{d}(v_d - v_{d+})R_{CX} \nonumber\\
     &+\partial_\parallel\left( D_d m_{d} n_{d} v_{d} \frac{\partial_\parallel p_d}{p_d} \right)
     + \partial_\parallel\left( D_d  m_d n_d \partial_\parallel T_d\right)
\end{align}
in terms of the momentum advection, the parallel pressure gradient, the electric field (for the ions only), momentum transfer between the species due to ionisation, recombination and charge exchange (where the charge exchange rate is $R_{CX}=k_{CX}n_d n_{d+}$), and enhanced diffusion (for neutrals only) as an approximation of cross-field transport.
The deuterium ion, neutral and electron temperature evolution is given by energy conservation equations (where the energy is related to pressures via $\varepsilon = \frac{3}{2}(p_e + p_i)=\frac{3}{2}n(T_e + T_i)$)
\begin{align}
    \frac{\partial}{\partial t}\left( \frac{3}{2} p_{d+} \right) =
    & v_{d+} \partial_\parallel p_{d+}
        -\partial_\parallel \left[ \frac{5}{2}p_{d+} v_{d+} - \kappa_{d+}\partial_\parallel T_{d+} \right] \nonumber\\
        &+(R_{CX} + R_{iz})\left(\frac{1}{2}m_d(v_d - v_{d+})^2\right) \nonumber\\
        &+R_{iz} T_d - R_{rec} T_{d+} + W_{d+,e} + S_{\varepsilon,d+}\\
       \frac{\partial}{\partial t}\left( \frac{3}{2} p_{d} \right) =
    & v_{d} \partial_\parallel p_{d}
        -\partial_\parallel \left[ \frac{5}{2}p_{d} v_{d} - \kappa_{d}\partial_\parallel T_{d} \right] \nonumber\\
        &+(R_{CX} + R_{rec})\left(\frac{1}{2}m_d(v_d - v_{d+})^2\right) \nonumber\\
        &+R_{rec} T_{d+} - R_{iz} T_d\nonumber\\
        &+\partial_\parallel\left( D_d \frac{3}{2} p_{d} \frac{\partial_\parallel p_d}{p_d} \right)
        + \partial_\parallel\left( \frac{5}{2} D_d n_d \partial_\parallel T_d\right)\\
        \frac{\partial}{\partial t}\left( \frac{3}{2} p_{e} \right) =
    & v_{e} \partial_\parallel p_{e}
        -\partial_\parallel \left[ \frac{5}{2}p_{e} v_{e} - \kappa_{e}\partial_\parallel T_{e} \right] \nonumber\\
        &+\varepsilon_{iz} R_{iz} + \varepsilon_{rec} R_{rec} -c_{Ne} n_e^2 L_{Ne}- W_{d+,e} + S_{\varepsilon,e}
\end{align}
in terms of the pressure advection and the parallel gradient of the convective and conducted heat fluxes. For the deuterium ions and neutrals, there are terms for the kinetic energy exchange due to charge-exchange and the ionisation/recombination, and for the pressure exchange due to ionisation and recombination. For the neutrals, there are additional diffusion terms to approximate cross-field transport. For the electrons, there are terms for the effective energy source or sink per ionisation and recombination event, and for the radiative power loss due to the neon impurity (with $L_z$ evaluated using OpenADAS data with $n_e\tau=0.5ms\cdot10^{20}m^{-3}$, from \cite{Henderson2023-hu}). For the electron and ions, there are also terms for the equipartition energy exchange $W_{d+,e}$ and external energy sources $S_{\varepsilon}$. For these simulations, we set equal electron and ion heat sources ($S_{\varepsilon,i} = S_{\varepsilon,e}$).
The equations are closed with no-flux boundary conditions at the `upstream' boundary -- which prevent fluxes of energy and particles through that boundary -- and with standard sheath boundary conditions at the `target' boundary \cite{Stangeby1995-qv}. We assume perfect recycling ($R=1$) at the target, and use a `pump' to remove neutrals from the divertor, where neutrals are assumed to be pumped with a finite residence time $\tau_{pump}= 100ms$\footnote{The order-of-magnitude for the particle residence time was estimated assuming roughly approximate pumping times for impurities and main ions, and using impurity concentration decay rates from Figure 5 of ref \cite{Henderson2024-bw}. This crude approximation of pumping should be extended in future work -- see i.e. ref\cite{Bosch1997-zf}.}.

\subsection{Numerical details}

The simulations are initialised with spatially flat profiles for each of the time-varying quantities, over a non-uniformly-spaced grid of 1000 points, with tighter packing towards the divertor target. Although it is possible to modify grid to model flux expansion in Hermes, this feature was not used for these simulations. The fields are evolved in time following the equation set introduced in section \ref{sec:hermes_equations}. Since we are mostly interested in the steady-state solution, we use the first-order-in-time-accurate backwards-Euler introduced in ref \cite{Dudson2024-ru}. Using this numerical method, quasi-steady-state conditions are reached in $10-100ms$, requiring a few hours on a single CPU. This is clearly not relevant for real-time modeling. Nevertheless, we are interested in this model for several reasons. The first is to identify the minimum physics description necessary to model detachment for a desired degree of accuracy: to develop, calibrate or train simpler models. Secondly, although this model is far from real-time, surrogate models have been developed for Hermes-3, such as in ref \cite{Holt2024-hi}. Therefore, identifying where this model should be extended to match higher fidelity modeling or experiments is valuable if simpler models cannot reach the desired degree of accuracy.

\section{Detachment in Hermes-1D}

\begin{figure}
    \centering
    \includegraphics[width=1\linewidth]{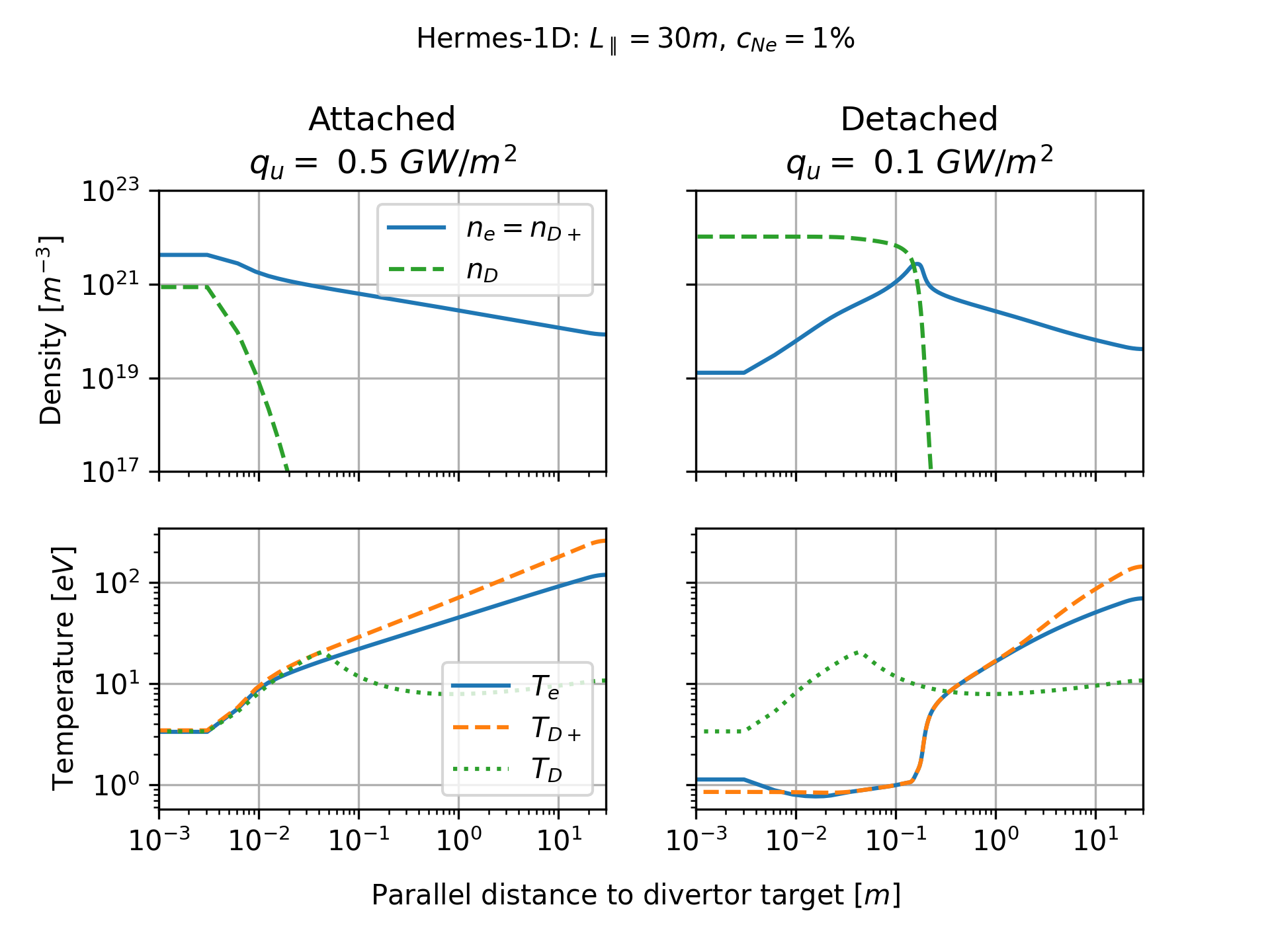}
    \caption{Two Hermes-1D simulations in steady state, showing an attached simulation (\textit{left column}) and a detached simulation (\textit{right column}). The \textit{top row} gives the electron and neutral-deuterium densities, and the \textit{bottom row} gives the electron, deuterium-ion and neutral-deuterium temperatures.}
    \label{fig:attached_versus_detached}
\end{figure}

\begin{figure}
    \centering
    \includegraphics[width=1\linewidth]{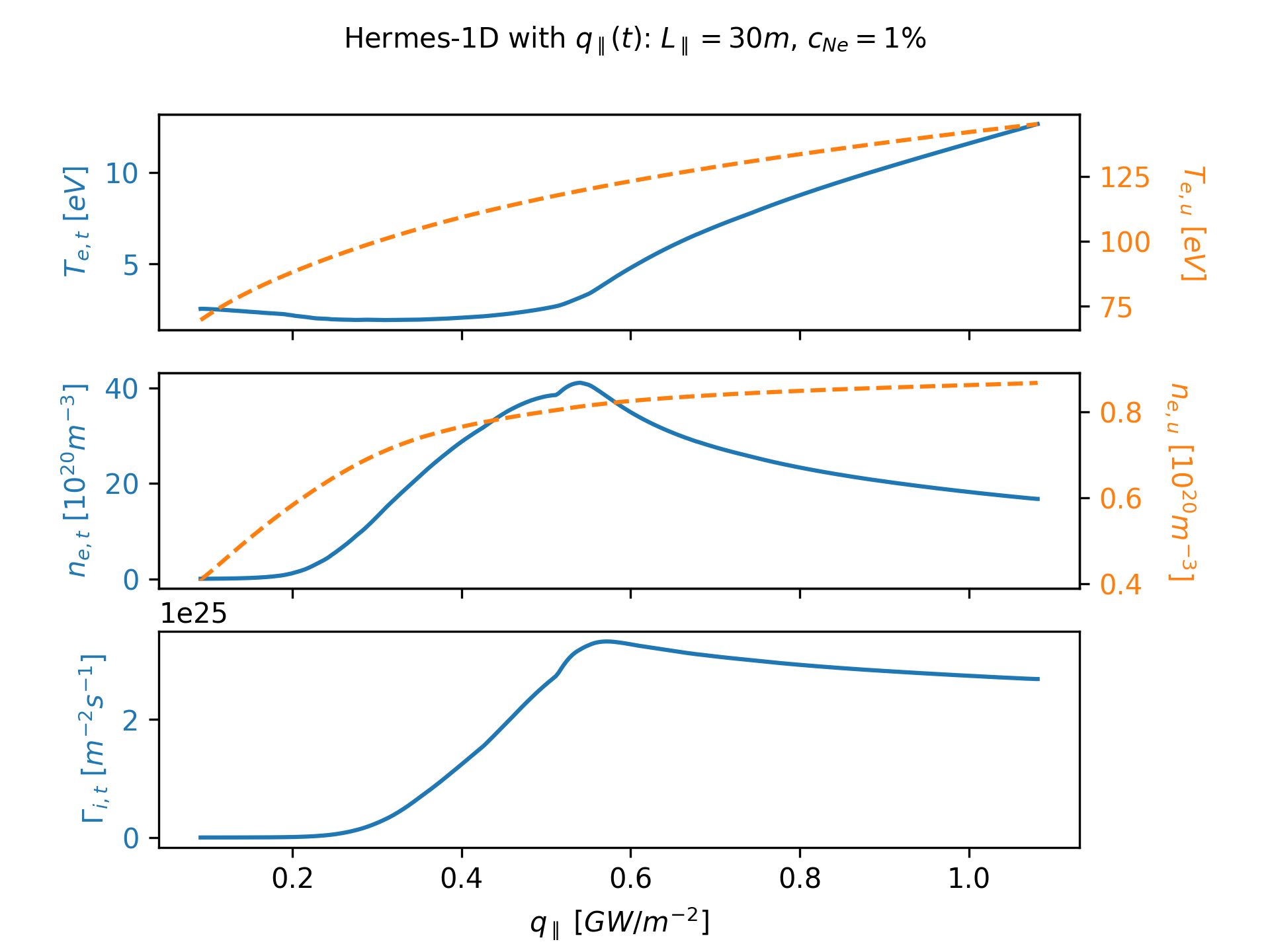}
    \caption{Target and upstream parameters from attached to detached conditions, giving the target (\textit{blue-solid, left axis}) and upstream (\textit{orange-dashed, right axis}) electron temperature (\textit{top row}), density (\textit{middle row}) and target particle flux (\textit{bottom row}) as a function of the input parallel heat flux density $q_\parallel$.}
    \label{fig:time_varying_source}
\end{figure}

As a first test of the Hermes-1D simulations, we perform simulations to demonstrate the transition from attached to detached plasma solutions by varying the input heat flux for a fixed impurity concentration and input particle flux. Distinctive 'attached' and 'detached' solutions are identified, as shown in figure \ref{fig:attached_versus_detached}. In attached conditions the neutral density is negligible throughout the entire domain, and the plasma temperature rises smoothly away from the divertor targets. In detached conditions, there is a sharp interface between two distinct regions — one where the neutral density exceeds the plasma density and the plasma temperature drops to $\sim 1eV$, and another upstream of this where the solution looks similar to the attached case. The neutral temperature remains high downstream of the detachment front, at approximately the same temperature as the ion temperature at the detachment front (where charge exchange is maximized), balancing the pressure upstream of the detachment front. 
We can investigate the attachment-detachment transition further by performing simulations with time-varying sources. In figure \ref{fig:time_varying_source} we sinusoidally varied the power from $0.1$ to $1.1 GW/m^2$ for a constant input particle flux rate. A period of $100ms$ was used for the power source variation to ensure that the instantaneous solution was always close to the steady-state solution.
We see that, as the power is decreased, the particle flux to the target initially increases due to the increasing target density. Then, at $0.4GW/m^2$, the target temperature levels out at $2eV$ and the target density begins to drop. This causes the target particle flux $\Gamma_{i,t}$ to roll over, which is a characteristic feature of detachment \cite{Krasheninnikov2017-ay}. For a fixed input particle flux and pumping residence time, we also observe that the upstream density drops as the input power flux decreases. This may be a sign of \textit{power starvation} \cite{Verhaegh2019-qc}, where the density drops as the power available for neutral ionisation is decreased. In our simulations, the particle loss rate due to pumping increases as the neutral density increases, and as such for a fixed particle source rate the total density in the flux tube must decrease.

\subsection{Hermes-1D detachment scalings}\label{sec:detachment_scalings}
\begin{figure}
    \centering
    \includegraphics[width=1\linewidth]{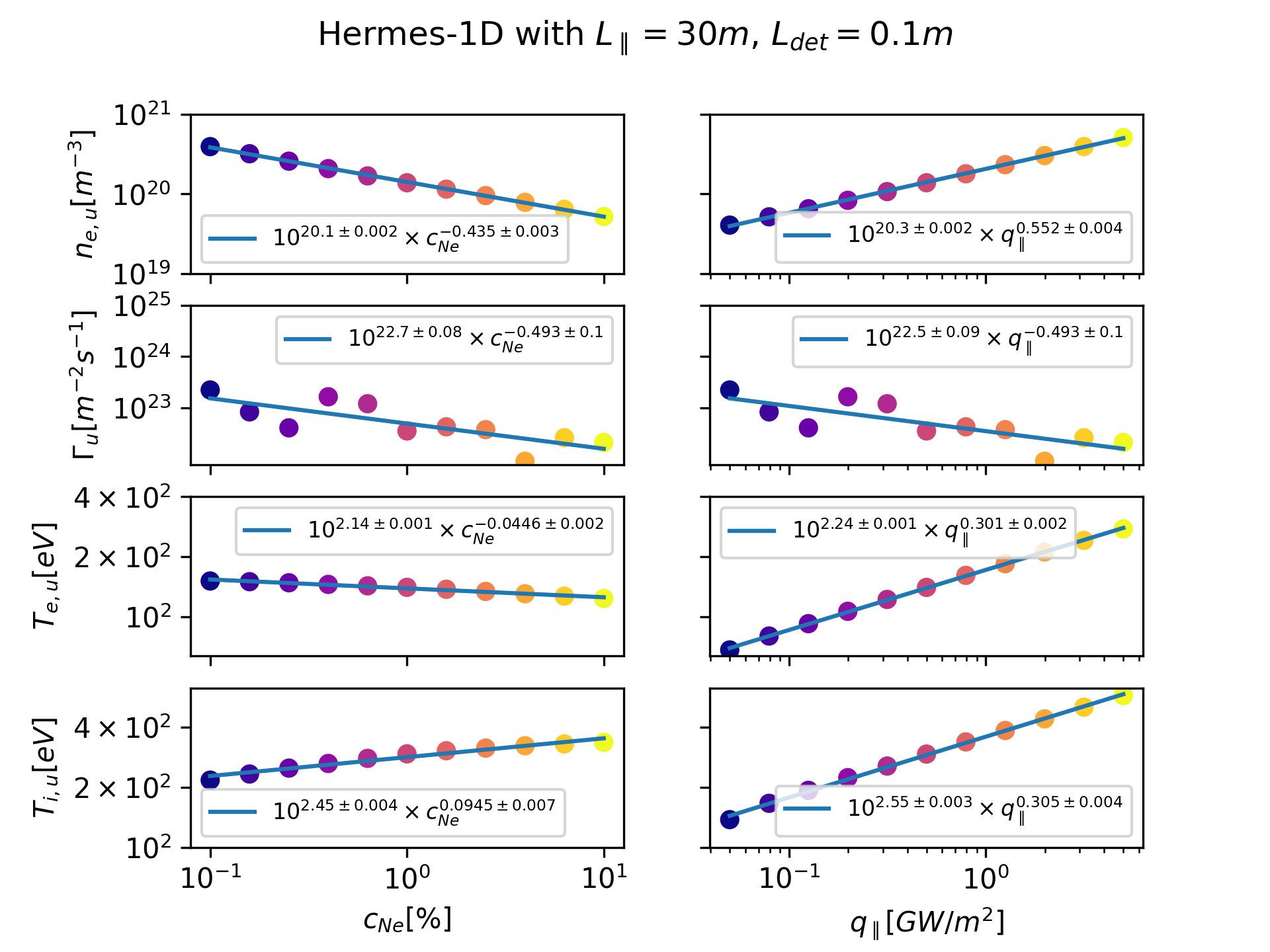}
    \caption{Upstream density (\textit{top row}), input particle flux (\textit{second row}), upstream electron temperature (\textit{third row}) and upstream ion temperature (\textit{bottom row}), as a function of the neon concentration (\textit{left column}) and input power flux (\textit{right column}). For the neon concentration scan, the power is held constant at $1GW/m^2$, and for the power flux scan the neon concentration is held constant at $1\%$. For each plot, simulation results are shown as colored points (with the color set to the x-value) and a power-law regression is shown as a solid line, with the regression given in the legend (in units given in the x- and y-labels).}
    \label{fig:detachment_scan}
\end{figure}
How does the impurity concentration $c_{Ne}$ relate to the input heat flux density $q_\parallel$ and the upstream density $n_u$ required for detachment in these simulations? We developed a simple PID controller to adjust the input particle flux $\Gamma_u$ or the input heat flux density $q_{\parallel,u}$ to control the position where the neutral density $n_d$ equals the electron density $n_e$. As seen in figure \ref{fig:attached_versus_detached}, this point corresponds to the point where the electron temperature drops sharply from $T_e\sim 10eV$ to $T_e\sim1eV$. For this work, we will refer to this point as the `detachment front'. Using the PID controller, we determined combinations of $c_{Ne}$, $q_{\parallel,u}$ and $n_u$ which resulted in steady-state detachment. In figure \ref{fig:detachment_scan} we show the results for 11 log-spaced values of $q_\parallel$ from $0.1$ to $10 GW/m^2$, and for 11 log-spaced values of $c_{Ne}$ from $0.1\%$ to $10\%$. For each simulation, the controller adjusts the input particle flux until the detachment front is stabilized at a parallel distance of $0.1m$ to the divertor target.
We then perform regressions of the results. We find that the simulation results are described by a power law, both for the impurity-concentration and heat flux scans. Combining the two, we find that
\begin{align}
    n_{e,u} &\simeq 10^{20.2\pm0.1} m^{-3} \left( \frac{c_{Ne}}{1\%} \right)^{-0.435\pm0.003} \left( \frac{q_\parallel}{1GW/m^2} \right)^{0.552\pm0.004}\label{eq:nu_scaling}\\
    \Gamma_{u} &\simeq 10^{22.6\pm0.1} \frac{m^{-2}}{s^{-1}} \left( \frac{c_{Ne}}{1\%} \right)^{-0.493\pm0.1} \left( \frac{q_\parallel}{1GW/m^2} \right)^{-0.493\pm0.1}\label{eq:gamma_scaling}\\
    T_{e,u} &\simeq 148\pm10  eV \left( \frac{c_{Ne}}{1\%} \right)^{-0.0446\pm0.002} \left( \frac{q_\parallel}{1GW/m^2} \right)^{0.301\pm0.002}\label{eq:Teu_scaling}\\
    T_{i,u} &\simeq 318\pm36 eV \left( \frac{c_{Ne}}{1\%} \right)^{0.0945\pm0.007} \left( \frac{q_\parallel}{1GW/m^2} \right)^{0.305\pm0.004}\label{eq:Tiu_scaling}
\end{align}
Of these, the upstream electron density $n_{e,u}$, the upstream electron temperature $T_{e,u}$ and the upstream ion temperature $T_{d+,u}$ closely follow power-law regressions. The upstream particle flux $\Gamma_u$ doesn't appear to follow a power-law. This is because, even though the detachment front is close to the desired position, the PID controller is still making small adjustments to stabilise the front at that position. The slight drop in $T_{e,u}$ and slight increase in $T_{d+,u}$ with increasing impurity concentration is due to the decrease in electron-ion thermal coupling as the upstream density decreases at high impurity concentrations.

\section{The Lengyel-Goedheer model}

The power-law scalings suggests that the detachment \textit{steady-state} may be described by simple models such as the one proposed by Lengyel \& Goedheer \cite{Lengyel1981-in}. Following the method of \cite{Moulton2021-id}, the Lengyel-Goedheer model is derived by assuming the parallel heat-flux-density is carried only by electron conduction, letting us write $q_\parallel = \kappa_\parallel T_e^{5/2} \frac{\partial T_e}{\partial s}$ for $s$ a coordinate along a field-line. If we add a fixed-fraction radiating impurity with a concentration of $c_z = n_z / n_e$, the change in $q_\parallel$ due to radiative cooling will be $\frac{\partial q}{\partial s}=n_e n_z L_z = n_e^2 c_z L_z$ where $L_z$ is the radiative cooling in $W m^{3}$ \cite{Mavrin2017-ey}. Under these assumptions
\begin{align}
q \frac{\partial q}{\partial s} = \kappa_\parallel T_e^{5/2} \frac{\partial T_e}{\partial s}n_e^2 c_z L_z(T_e)    
\end{align}
Integrating along the fieldline from the target $t$ to the upstream outboard-midplane $u$, we can write $\int_t^u q \partial q = \int_t^u \kappa_\parallel T_e^{5/2} n_e^2 c_z L_z(T_e) \partial T_e$. If we assume that the static electron pressure is constant along most of the fieldline $n_e T_e =n_{e,u}T_{e,u}$
\begin{align}\frac{1}{2}(q_u^2 - q_t^2) = \kappa_\parallel n_{e,u}^2T_{e,u}^2 c_z \int_t^u L_z(T_e) \sqrt{T_e} \partial T_e\end{align}
As a shorthand, we denote $\int_t^u L_z(T_e) \sqrt{T_e} \partial T_e = L_{INT}$ and write the impurity concentration required as
\begin{align}c_z = \frac{q_u^2 - q_t^2}{2 \kappa_\parallel n_{e,u}^2T_{e,u}^2 L_{INT}}\end{align}
The upstream density $n_{e,u}$ is a model input, while the upstream temperature $T_{e,u}$ can be determined in two ways. The first is via Spitzer-Harm power balancing $T_{e,u} = \left( T_{e,t} + \frac{7}{2}\frac{q_u L_\parallel}{\kappa_\parallel} \right)^{2/7}$  — which we refer to as the "basic Lengyel-Goedheer model". The second is by adjusting the upstream temperature so that it is consistent with the heat flux profile. This is done by rearranging and integrating $q_\parallel = \kappa_\parallel T_e^{5/2} \frac{\partial T_e}{\partial s}$, giving $L_\parallel = \kappa_\parallel \int_t^u T_e^{5/2} / q_\parallel(T_e) \partial T_e$ with $q_\parallel(T_e) = q_t^2 + 2\kappa_\parallel n_{e,u}^2T_{e,u}^2 c_z \int_t^{T_e} L_z(T_e') \sqrt{T_e'} \partial T_e'$. We iterative solve this for a consistent $(T_{e,u}, c_z)$ pair until $L_\parallel$ matches the actual connection length \cite{Reinke2017-qx}  — which we refer to as the "full Lengyel-Goedheer model".
We can further simplify the basic Lengyel-Goedheer model by assuming $T_{e,t}\to 0, q_t \to 0$. Using $L_{INT}=m_L(Z, n_e\tau)T_{e,u}$ from ref \cite{Reinke2017-qx} and $T_u\propto q_u^{2/7}$ from Spitzer-Harm power balancing gives
\begin{align}
    c_z \propto q_u^2 \left( n_{e,u}^{2} T_{e,u}^{3} \right)^{-1}\propto q_u^{8/7}n_{e,u}^{-2}\label{eq:reduced_lengyel}
\end{align}
As an aside, this is sometimes converted to a scaling in terms of the power crossing the separatrix, by assuming $\lambda_q\sim B^{-1}$ \cite{Eich2013-yg} which lets us write $q\sim \frac{PB}{R}$. Ignoring the extra factor of $1/7$ on $q_u$, we can write $c_z \sim \left(\frac{PB}{R}\right)/n_{sep}^2$ which is often used as a simple metric for detachment accessibility.

\subsection{Comparing the Hermes and Lengyel-Goedheer model}

In the Lengyel-Goedheer (with Reinke's $L_z \propto T_{e,u}$ approximation, equation \ref{eq:reduced_lengyel}) the impurity concentration required to reach detachment scales as $c_z \propto q_u^{8/7}n_{e,u}^{-2}\sim q_u^{1.14}n_{e,u}^{-2}$. Rearranging regression \ref{eq:nu_scaling} for $c_{Ne}$, we find that $c_{Ne}\propto q_\parallel^{1.269} n_{e,u}^{-2.30}$ which agrees reasonably well with the simple scaling. We can also compare regressions \ref{eq:Teu_scaling} and \ref{eq:Tiu_scaling} to the Spitzer-Harm estimate for $T_{e,u} = \left( T_{e,t}^{7/2} + 3.5 \frac{q_{\parallel,u} L_\parallel}{\kappa_e}\right)^{2/7}\propto q_{\parallel,u}^{2/7\sim 0.286}$. For the electron temperature, the calculated regression has a slightly stronger dependence on $q_{\parallel,u}$ than the Spitzer-Harm model ($0.301$ instead of $0.286$). The ion temperature follows a similar scaling to the Spitzer-Harm scaling, with $T_{i,u}\propto q_{\parallel,u}^{0.305}$ — possibly as a consequence of strong coupling of the electron and ion temperatures close to the divertor targets. If we instead use the $T_{e,u}$ regression \ref{eq:Teu_scaling} (keeping the weak $c_z$ dependence) in $c_z \propto q_u^2 \left( n_{e,u}^{2} T_{e,u}^{3} \right)^{-1}$ we find $(c_z\propto q_\parallel^{1.24}n_{e,u}^{-2.27})$, giving an almost exact match between the simplified Lengyel-Goedheer model and regression \ref{eq:nu_scaling}.

\begin{figure}
    \centering
    \includegraphics[width=1\linewidth]{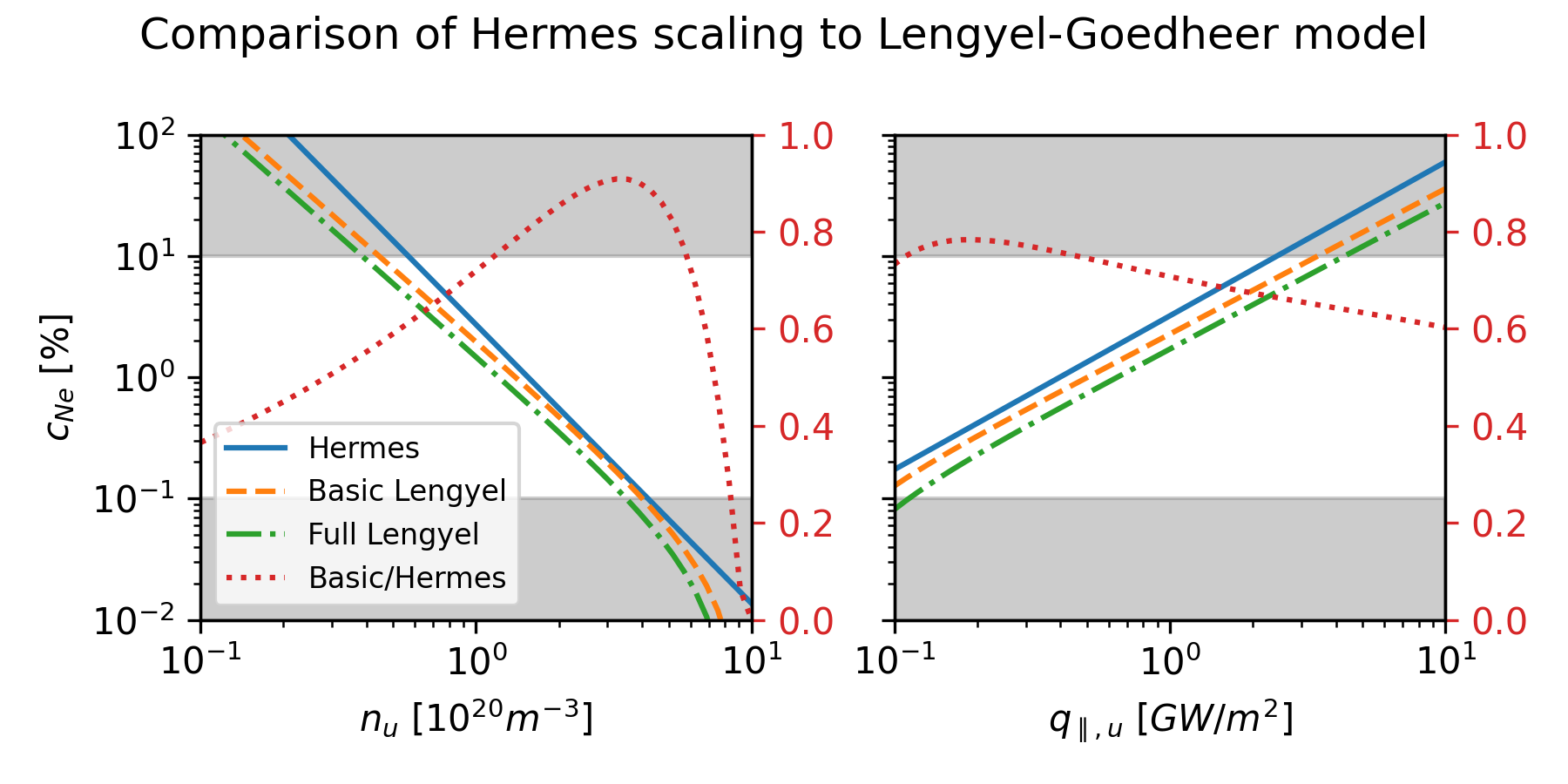}
    \caption{Neon concentration as a function of the upstream density (\textit{left}) and input power flux (\textit{right}). For the neon concentration scan, the power is held constant at $1GW/m^2$, and for the power flux scan the neon concentration is held constant at $1\%$. For each scan, the scaling derived from Hermes is shown as a \textit{blue, solid} line. This is compared to results from the basic Lengyel-Goedheer model, shown as an \textit{orange, dashed} line, and to the full Lengyel-Goedheer model, shown as a \textit{green, dot-dashed} line. The ratio of the Hermes result to the basic Lengyel-Goedheer model is shown as a \textit{red, dotted} line, and should be compared to the right-hand \textit{red} axis. The regions when the Hermes regressions are extrapolated are indicated by the \textit{gray background}.}
    \label{fig:hermes_versus_lengyel}
\end{figure}

In addition to parameter scalings, the Lengyel-Goedheer model can also be used to predict the absolute concentration of the impurity concentration required to reach detachment. We implemented the Lengyel-Goedheer model in the open-source \href{https://github.com/cfs-energy/cfspopcon}{cfsPOPCON} framework \cite{Body2024-mk} and compared the results to the scalings derived from Hermes. In figure \ref{fig:hermes_versus_lengyel} we see that Hermes and the Lengyel-Goedheer model predict similar parameter scalings (as expected) and absolute values within a factor of 2 for the region which we scanned with Hermes.
Hermes predicts a slightly lower impurity concentration required for detachment than the Lengyel-Goedheer model, in contrast to recent comparisons of the Lengyel-Goedheer model to SOLPS 2D transport simulations. In Moulton et al., 2021 \cite{Moulton2021-id}, the Lengyel-Goedheer model was found to estimate a $4.3\times$ higher impurity concentration than SOLPS-4.3 for neon-seeded ITER conditions, and in Jarvinen et al., 2023 \cite{Jarvinen2023-xh}, the Lengyel-Goedheer model was found to estimate a factor of between $5$ and $10\times$ higher impurity concentration than SOLPS-ITER for argon-seeded DEMO-ADC conditions. In both studies, the largest contribution to the deviation was the assumption of electron-conduction-dominated heat transport in the Lengyel-Goedheer model. In the Hermes simulations, by contrast, we find that the parallel heat transport upstream of the detachment front is dominated by conduction (typically by a factor of $-\kappa_e\partial_\parallel T_e / (5/2 p_{d+} v_{d+})\sim100$, while downstream of the detachment front the convective heat flux dominates). The reduced impact of convection is likely due to our treatment of neutrals in our Hermes simulations. We have assumed that neutral transport is confined to our flux-tube — effectively equivalent to assuming extremely high neutral baffling. This has been shown in SOLPS-ITER simulations to decrease parallel convection relative to a more open divertor \cite{Cowley2024-ns}. To increase the contribution of heat convection and bring the Hermes results closer to SOLPS-ITER we need to introduce transport channels for neutrals outside of the simulated flux-tube — such as through the use of 'neutral reservoirs' \cite{Derks2022-ug,Derks2024-nj}. In addition to parallel convection, cross-field plasma transport (i.e. divertor broadening) was also found to be important in the SOLPS simulations as the simulations approached detachment \cite{Moulton2021-id} — which could be treated as an effective flux expansion in 1D simulations \cite{Derks2022-ug,Derks2024-nj}. These extensions are currently being implemented in Hermes, and we will investigate their impact in future work.

\subsection{A look ahead: time-dependence}
\begin{figure}
    \centering
    \includegraphics[width=1\linewidth]{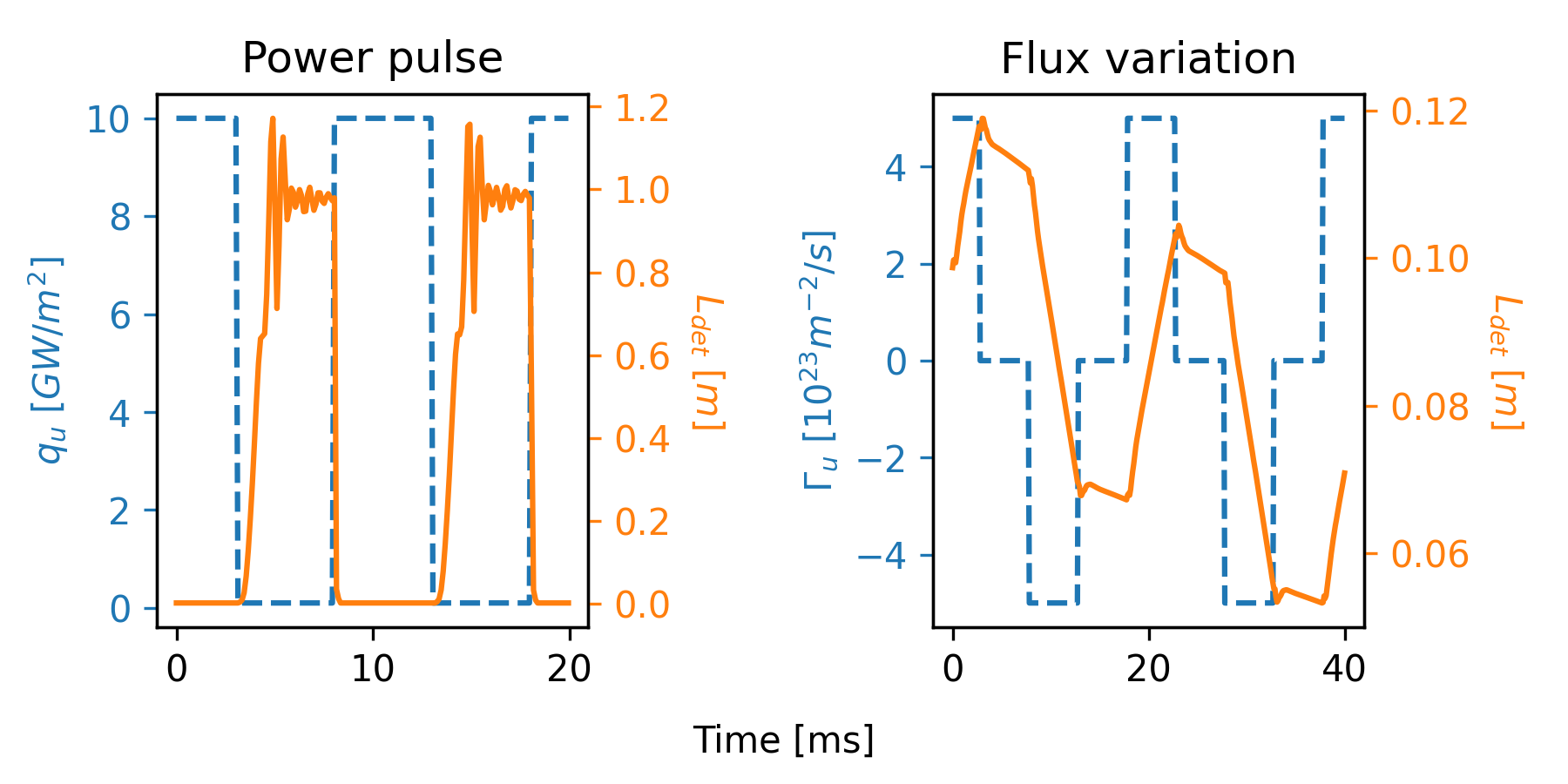}
    \caption{Response of the detachment front location (shown as a \textit{solid orange line}) to a sharply-varying source (shown as a \textit{dashed blue line}). The \textit{left figure} shows the response to a sharply-varying heat flux source, and the \textit{right figure} shows the response to a sharply-varying particle flux source.}
    \label{fig:shock_response}
\end{figure}
The steady-state solution of the 1D modeling presented in this paper can largely be described by the Lengyel-Goedheer model. This indicates that, if the plasma reaches steady-state on a faster timescale than we can detect or react, models like the Lengyel-Goedheer model may be suitable for control. As a first look ahead to the timescales of detachment, we performed simulations where we varied the input heat flux $q_\parallel$ and the input particle source $\Gamma_u$. Instead of using slowly varying sources to keep the solution close to steady-state, we used Heaviside functions to test how the system responds to perturbations faster than the equilibration time. In figure \ref{fig:shock_response}, we show the results of a heat flux source sharply varying from $0.1-10GW/m^2$, and of a particle flux source sharply varying between $ 5, 0,\text{and}-5\times10^{23}m^{-2}/s$ (i.e. we apply a particle \textit{sink} upstream when the source is negative). We stress that these values and timescales are \textit{not representative} of a physically-realistic system, but rather intended as a demonstration.
The divertor very quickly reattaches when the heat flux is increased (in $\sim0.1ms$), and then takes longer ($\sim2ms$) to recover to its original position once the heat flux is decreased. Conversely, the detachment front doesn't reach the divertor target when we cut the gas fueling, nor when we start removing particles upstream. This appears to be a consequence of gas fueling model — reducing the neutral residence time ($\tau_{pump}$ in equation \ref{eq:neutral_continuity}) by a factor of 10 increases the velocity of the detachment front accordingly (from $0.6$ to $6m/s$). We also see that the detachment front doesn't return to its original position after we start fueling gas again, but rather the cycle-averaged detachment front position drifts at the zero-fueling velocity (since the cycle-average fueling rate is zero). Although the perturbation magnitude and timescales are unphysical (leading to unphysical system response times), it seems reasonable that the detachment front responds more quickly to a heat pulse (which is transported to the detachment front via heat conduction) than to a particle-flux pulse (which is transported via slower advection). A steady-state model would be a useful description of the heat flux pulse shown in figure \ref{fig:shock_response}, since the system equilibrates faster than we would be able to respond. Conversely, for the particle flux pulse, a time-dependent model would be useful, since the system state depends on its history as well as on the input parameters. In future work, we will explore the impact of time-dependence, improving the accuracy of the pumping model and using more realistic pulse shapes and timescales, comparing to experimental results such as the reattachment times reported in ref \cite{Henderson2024-bw}.

\section{Conclusion}

Reduced scrape-off-layer models are needed for detachment control on future fusion devices such as SPARC and ARC. In this work, we used the configurable Hermes framework to investigate detachment in a simple, fixed-fraction-impurity 1D Braginskii model. We show that the model is able to reproduce both attached and detached cases, as well as the transition between the two states and the rollover of the ion saturation current. We then implemented a PID controller for the detachment position which holds the detachment front at a desired position. This enabled scans to determine combinations of the input heat flux, upstream density and impurity concentration which led to detachment. The results of these scans are described by simple power-laws, which we compare to the simpler Lengyel-Goedheer model. We show that the Lengyel-Goedheer model agrees closely with the scalings derived from Hermes, both in terms of magnitude and parametric inter-dependencies. Since previous work has shown that the Lengyel-Goedheer scaling matches SOLPS simulations reasonably well, this gives us confidence that the Hermes simulations are also reasonably accurate. However, unlike the SOLPS simulations, the Hermes simulations predict almost exactly the same values as the Lengyel-Goedheer -- potentially due to the very high effective baffling or the lack of cross-field transport. We finally demonstrate how time-dependence may become important for modeling and controlling slowly-varying systems, and propose a series of extensions for Hermes towards the development of accurate, time-dependent simulations.

\section{Data availability}

The software packages used for this paper are all publicly available. The results in this paper were generated using the following software versions (given as \texttt{software name}: git hash, link to software repository).
\begin{itemize}
\item \texttt{Hermes-3}: \href{https://github.com/bendudson/hermes-3/commit/8c450228e61d66200de225c168aa80fdfd91f7b2}{8c45022} available from \href{https://github.com/bendudson/hermes-3}{github.com/bendudson/hermes-3}
\item \texttt{cfsPOPCON}: \href{https://github.com/cfs-energy/cfspopcon/commit/88e077fccb2483b3d4b389a15e74f61dd284a9e7}{88e077f} available from \href{https://github.com/cfs-energy/cfspopcon}{github.com/cfs-energy/cfspopcon}
\item \texttt{radas}: \href{https://github.com/cfs-energy/radas/commit/92812df646c09c115c08bcfcb47853dff13f8950}{92812df} available from \href{https://github.com/cfs-energy/radas}{github.com/cfs-energy/radas}
\end{itemize}

Input files and simulation results for this paper are available at \href{https://doi.org/10.5281/zenodo.11484706}{10.5281/zenodo.11484706}.

\section{Acknowledgements}
This work was supported by Commonwealth Fusion Systems. Prepared in part by LLNL under Contract DE-AC52-07NA27344.

\section{References}

\printbibliography

\end{document}